\documentclass[aps,prl,reprint,preprintnumbers,superscriptaddress,amsmath,amssymb]{revtex4-2}

\usepackage{graphicx}
\graphicspath{{./}{../../}}

\begin{document}

\preprint{KEK-TH-2862}
\preprint{RIKEN-iTHEMS-Report-26}

\title{A Rotor-Dressed Semiton State in a Monopole--Fermion Model}

\author{Yuta Hamada}
\affiliation{Theory Center, IPNS, High Energy Accelerator Research Organization (KEK), 1-1 Oho, Tsukuba, Ibaraki 305-0801, Japan}
\affiliation{Graduate Institute for Advanced Studies, SOKENDAI, 1-1 Oho, Tsukuba, Ibaraki 305-0801, Japan}
\affiliation{RIKEN Center for Interdisciplinary Theoretical and Mathematical Sciences (iTHEMS), RIKEN, Wako 351-0198, Japan}

\begin{abstract}
In four-flavor massless QED with a magnetic monopole, a single incident fermion is scattered into a state carrying a half-integral current vector in each flavor channel, called the semiton.
Recently, the semiton has been interpreted as a twisted-sector fermion, but its microscopic Hilbert-space representation has not been fully developed.
In this Letter, using a bosonized fermion--rotor model, we construct the semiton state explicitly.
This is achieved by first identifying the correct vacuum of the system, and then constructing a unitary operator that shifts the rotor and creates a particle-hole cloud in each flavor channel.
The semiton state is obtained by applying this operator to an ordinary fermion state.
The resulting state has a localized fermion flavor density in the core region and a half-integral semiton front at the wavepacket. We also show that the state carries the correct $U(1)_M$ charge expectation values and energy-density profile.
Our result provides a concrete Hilbert-space realization of the varying-Fock-space interpretation of monopole--fermion scattering, advocated in previous works by the author and collaborators.
\end{abstract}

\maketitle

\paragraph{Introduction.---}\label{introduction}

It is known that the scattering between a magnetic monopole~\cite{tHooft:1974kcl,Polyakov:1974ek} and a charged massless fermion leads to violation of baryon number conservation \cite{Rubakov:1982fp,Callan:1982au,Callan:1982ah,Callan:1982ac}. 
When the number of flavors is four, corresponding to one generation in the Grand Unified Theory (GUT), Callan found that the scattering of a single fermion in one channel leads to a half-integer current in each channel \cite{Callan:1983tm}.
This half-integer charge state is called the semiton, whose physical interpretation has been intensively discussed in the literature \cite{Polchinski:1984uw,AffleckSagi:1993bcft,Yegulalp:1994eq,Maldacena:1995pq,Kitano:2021pwt,Csaki:2021ozp,Brennan:2021ewu,Hamada:2022eiv,vanBeest:2023dbu,Khoze:2023kiu,Brennan:2023tae,vanBeest:2023mbs,Loladze:2024ayk,Loladze:2025jsq,Tachikawa:2026cxd}.

One promising interpretation of the semiton is as a fermion in a twisted sector, distinct from the ordinary Fock space~\cite{Maldacena:1995pq,Smith:2020nuf,Hamada:2022eiv,vanBeest:2023dbu}. Indeed, we can show that the twisted-sector fermion has the same quantum numbers as the semiton.
Nevertheless, it remains unclear why the twisted-sector fermion appears in low-energy scattering, given that no twisted boundary condition is imposed in the original GUT model.

A hint toward understanding the origin of the twisted sector was given in our previous work \cite{Hamada:2022eiv}.
There, we started from Polchinski's fermion--rotor model, which retains the relevant degrees of freedom of the monopole--fermion system.
By solving the Heisenberg equations, we showed that the rotor degree of freedom $\alpha$ changes during the scattering, and this change of $\alpha$ can be viewed as a change in the effective boundary condition.

Building on this observation, in this Letter, we explicitly construct a semiton state in the Hilbert space of the fermion--rotor model using the bosonized language.
A key ingredient is the entanglement between the rotor and the Dirac sea of the fermion in the vacuum state.
If we assume that the rotor degree of freedom is heavy, and the vacuum state is just a tensor product,
\begin{equation}
|\Omega\rangle_{\rm rot}\otimes|\Omega\rangle_{\rm fermion},\notag
\end{equation}
then there is no room to realize the semiton state in the Hilbert space.
However, we will show that the vacuum state is entangled, and consequently, the untwisted and twisted fermion states are realized as different patterns of entanglement.

More concretely, after identifying the vacuum state, the ordinary fermion state is realized by acting with a fermion vertex operator on the vacuum.
Then, we construct an operator that translates the rotor by half a period while coherently displacing the bosonized current profile of nonzero modes. 
The semiton state is then obtained by acting with this operator on the ordinary fermion state.
Its fermion number density divides into a core cloud and the half-integral semiton front.
We then check the validity of the construction by computing its energy density and its infrared overlap with an undressed fermion.

\paragraph{Fermion--rotor Hamiltonian and vacuum.---}\label{fermionrotor-hamiltonian-and-the-branch-local-vacuum}

We consider the setup where the monopole is located at the origin, and perform the $s$-wave reduction which is valid for low-energy scattering~\cite{Kazama:1976fm}.
In this way, the $(3+1)$-dimensional problem is reduced to a $(1+1)$-dimensional one.
We denote the left- and right-moving fermions by $\psi_i^L$ and $\psi_i^R$, respectively, where $i=1,\cdots,4$ is the flavor index.
In the unfolded picture, the left- and right-moving fermions are defined on the intervals $x\in[-L,0]$ and $x\in[0,L]$, respectively, where $L$ is the size of the system.
The two endpoints $x=\pm L$ are identified through the boundary condition $\psi_i^R(L)=-\psi_i^L(-L)$\footnote{To be precise, this boundary condition breaks the $U(1)_M$ symmetry introduced below. We will discuss this point later.}.
For $-r_0\leq x\leq r_0$, the fermions are coupled to the monopole rotor, whose coordinate and conjugate momentum are denoted by $\alpha\sim\alpha+2\pi$ and $\Pi$, respectively.
The moment of inertia of the rotor is denoted by $I$.

There are $U(1)_M$ and $SU(4)_{\rm f}$ transformations inherited from the $U(1)$ gauge and $SU(4)$ flavor symmetries of the original $(3+1)$-dimensional massless QED. 
The representations of the fermions are summarized in Table~\ref{tab:chiral-field-charges}.
As we can see in the first two rows, an incoming $\psi_i^L$ fermion cannot simply turn into $\psi_i^R$ after the scattering because the $U(1)_M$ charge would be reversed.
Consequently, the outgoing state must be the semiton in the fourth row, whose description in the Hilbert space has remained unclear. The goal of this Letter is to provide such a realization.

\begin{table*}[t]
\squeezetable
\caption{$U(1)_M$ weights and $SU(4)_{\rm f}$ representations.
The flavor vector is read from the first column.
For example, the outgoing semiton state has a flavor vector $\mathbf e_a-\frac{1}{2}\sum_i\mathbf e_i$.
}
\label{tab:chiral-field-charges}
\centering
\begin{ruledtabular}
\begin{tabular}{ccc}
Multiplet or state & $Q_M$ & $SU(4)_{\rm f}$\\
\colrule
$\psi_i^L$ & $+1$ & $\mathbf4$\\
$\psi_i^R$ & $-1$ & $\mathbf4$\\
Incoming semiton $\psi_a^L-\frac{1}{2}\sum_i\psi_i^L$ & $-1$ & $\mathbf4$\\
Outgoing semiton $\psi_a^R-\frac{1}{2}\sum_i\psi_i^R$ & $+1$ & $\mathbf4$\\
\end{tabular}
\end{ruledtabular}
\end{table*}

To this end, we introduce the Hamiltonian of the bosonized fermion--rotor system of Ref.~\cite{Polchinski:1984uw} (the bosonized form is obtained from Eq.~(2.1) of Ref.~\cite{Yegulalp:1994eq} as detailed in Sec.~S1 of the Supplemental Material~\cite{SupplementalMaterial}):
\begin{align}
\begin{aligned}
H_{\rm cur}^{(M)}
&=\frac{\Pi^2}{2I}
+\pi\sum_{i=1}^{4}\int_{-L}^{L}dx\,
{:}\bigl[J_{i,M}^{\rm cov}(x)\bigr]^2{:} ,
\end{aligned}
\label{eq:regulated-current-hamiltonian}
\end{align}
where
\begin{align}
\rho_{i,M}^{\rm can}(x)
&=\frac{F_i^{\rm can}}{2L}
+\frac{1}{2L}\sum_{n=1}^{M}\sqrt n\left(
b_{i,n}e^{ip_nx}+b_{i,n}^{\dagger}e^{-ip_nx}
\right),
\label{eq:canonical-density-expansion}\\
J_{i,M}^{\rm cov}(x)
&=\rho_{i,M}^{\rm can}(x)-\frac{\alpha}{2\pi}f(x).
\label{eq:covariant-current-definition}
\end{align}
For the bosonized currents, we keep the modes $p_n=\pi n/L$, $1\le n\le M$, and $F_i^{\rm can}$ is the zero mode corresponding to the fermion number of each flavor.
The annihilation and creation modes of the bosonized current are denoted by $b_{i,n}$ and $b_{i,n}^{\dagger}$, and normalized as $[b_{i,n},b_{j,m}^{\dagger}]=\delta_{ij}\delta_{nm}$.
These oscillator modes carry zero total fermion number and represent superpositions of particle--hole pairs.
The function $f(x)$ is a real, even core profile supported on $[-r_0,r_0]$ and normalized by $\int_{-r_0}^{r_0}dx\,f(x)=1$.
In every expression regularized at finite $M$, the same symbol $f$ denotes the corresponding regularized function. This remark applies to other functions we will introduce below, such as $g$ and $h$.
Normal ordering is with respect to $b_{i,n}$ and
$b_{i,n}^{\dagger}$.
The physical cutoff $\Lambda$ corresponds to $\Lambda=\pi M/L$.

The Hilbert space associated with the currents is
\begin{align}
\begin{aligned}
\mathcal H_{\rm cur}^{(M)}
&=\ell^2(\mathbb Z^4)_{\rm zero\ mode}
\otimes\mathcal H_{{\rm osc},M},
\end{aligned}
\label{eq:current-hilbert-space}
\end{align}
where $\mathcal H_{{\rm osc},M}$ is the Fock space of $b$ and $b^\dagger$, and $\ell^2(\mathbb Z^4)_{\rm zero\ mode}$ is the Hilbert space for the fermion flavor number $F_i^{\rm can}$.
The space $\ell^2(\mathbb Z^4)_{\rm zero\ mode}$ is spanned by $|\mathbf n\rangle$ with $n_i\in\mathbb Z$, and $F_i^{\rm can}$ acts on it as $F_i^{\rm can}|\mathbf n\rangle=n_i|\mathbf n\rangle$.
In addition to $\mathcal H_{\rm cur}^{(M)}$, we have a rotor Hilbert space spanned by $|\alpha\rangle_{\rm rot}$.

Since only the flavor-singlet current, $b_{\parallel,n}=\frac{1}{2}\sum_i b_{i,n}$, couples to the rotor, $H_{\rm cur}^{(M)}$ decomposes as
\begin{align}
H_{\rm cur}^{(M)}=H_{\rm zm}+H_{\parallel+{\rm rot}}^{(M)}+H_{\perp}^{(M)},
\end{align}
where $H_{\rm zm}=\frac{\pi}{2L}\sum_i(F_i^{\rm can})^2-\frac{\alpha}{2L}\sum_iF_i^{\rm can}$ and $H_{\perp}^{(M)}$ is the free Hamiltonian for the flavor non-singlet orthogonal currents.
The form of $H_{\parallel+{\rm rot}}^{(M)}$ will be shown below.

The perfect square structure of the Hamiltonian~\eqref{eq:regulated-current-hamiltonian} indicates that the system possesses a multiple-branch structure (see the End Matter for details).
For the purposes of this Letter, however, it suffices to focus on one branch, $\mathbf F^{\rm can}=\mathbf0$, and to lift $\alpha$ locally to $\mathbb R$.
By introducing
\begin{align}
x_n=\sqrt{\frac{L}{2\pi n}}
(b_{\parallel,n}+b_{\parallel,n}^{\dagger}),
\qquad
\pi_n=-i\sqrt{\frac{\pi n}{2L}}
(b_{\parallel,n}-b_{\parallel,n}^{\dagger}),
\label{eq:singlet-current-canonical-coordinates}
\end{align}
with $n=1,\ldots,M$, and the vectors $z=(\alpha,x_1,\ldots,x_M)^T$ and
$\boldsymbol\pi=(\Pi,\pi_1,\ldots,\pi_M)^T$, the interacting part of the Hamiltonian is\footnote{The constant term arises from the commutation relation between $x_n$ and $\pi_n$.}

\begin{align}
\begin{aligned}
H_{\parallel+{\rm rot}}^{(M)}
&=\frac{1}{2}\boldsymbol\pi^TA\boldsymbol\pi
+\frac{1}{2}z^TBz+{\rm const.},
\\
A&={\rm diag}(I^{-1},1,\ldots,1).
\end{aligned}
\label{eq:quadratic-rotor-current-hamiltonian}
\end{align}

\begin{align}
\begin{aligned}
B&=
\begin{pmatrix}
C&-\mathbf c^T\\
-\mathbf c&B_{xx}
\end{pmatrix},
\quad
(B_{xx})_{nm}=p_n^2\delta_{nm},\\
c_n&=\frac{\sqrt{2\pi}\,n f_n}{L^{3/2}},
\quad
C
=\frac{1+2\sum_{n=1}^{M}f_n^2}{\pi L} .
\end{aligned}
\label{eq:rotor-current-potential-matrix}
\end{align}

Here $f_n=\int_{-L}^{L}dx\,f(x)e^{i\pi nx/L}$.
The ground-state wavefunction is then

\begin{align}
\Psi_{\rm loc}(\alpha,\mathbf x)
=\left(\frac{\det K}{\pi^{M+1}}\right)^{1/4}
\exp\!\left[-\frac{1}{2}z^TKz\right],
\label{eq:branch-local-gaussian-wavefunction}
\end{align}
where
\begin{align}
K=A^{-1/2}
\left(A^{1/2}BA^{1/2}\right)^{1/2}A^{-1/2},
\qquad KAK=B .
\label{eq:gaussian-kernel-matrix}
\end{align}
Since $A$ is diagonal, $A^{-1/2}$ is just an inverse square root of each element of $A$.
Regarding $\left(A^{1/2}BA^{1/2}\right)^{1/2}$, we take the positive square root of the eigenvalues of $A^{1/2}BA^{1/2}$.
Note that the exponent of the wavefunction is written as
\begin{align}
\begin{aligned}
z^TKz
&=(\mathbf x+\alpha\boldsymbol\eta_K)^TK_{xx}
(\mathbf x+\alpha\boldsymbol\eta_K)+\Gamma\alpha^2,\\
\boldsymbol\eta_K&=K_{xx}^{-1}K_{x\alpha},\\
\Gamma&=K_{\alpha\alpha}
-K_{\alpha x}K_{xx}^{-1}K_{x\alpha}>0 .
\end{aligned}
\label{eq:gaussian-schur-completion}
\end{align}
Equivalently, the vacuum state is
\begin{align}
\begin{aligned}
|\Omega_{\mathbf0}^{(M)}\rangle
&=\int_{\mathbb R}d\alpha\,
\left(\frac{\Gamma}{\pi}\right)^{1/4}e^{-\Gamma\alpha^2/2}
|\alpha\rangle_{\rm rot}\\
&\quad\otimes|\mathbf0\rangle_{\rm zero}
\otimes|\Omega_{\parallel}^{(M)}(\alpha)\rangle
\otimes|0\rangle_\perp .
\end{aligned}
\label{eq:branch-local-vacuum-ket}
\end{align}
where $|\mathbf0\rangle_{\rm zero}\in\ell^2(\mathbb Z^4)_{\rm zero\ mode}$, $|0\rangle_\perp$ is the vacuum state of $H_\perp^{(M)}$, and $\Psi_{\rm loc}$ is the $(\alpha,\mathbf x)$ representation of the state $|\Omega_{\mathbf0}^{(M)}\rangle$.
Since $\boldsymbol\eta_K\ne0$,
Eq.~\eqref{eq:branch-local-vacuum-ket} is not a product of a rotor wavefunction and a fixed Dirac sea.
Changing the rotor coordinate changes the fermionic vacuum itself.
More quantitatively, at fixed physical cutoff $\Lambda$, the rotor--Dirac-sea entanglement entropy is nonzero.

\paragraph{Wavepacket vertex and joint-displacement algebra.---}\label{wavepacket-vertex-and-joint-displacement-algebra}
Using the standard bosonized fermion vertex $\mathcal V_{a,M}^{\dagger}(x)$~\cite{vonDelftSchoeller:1998bosonization} (see also End Matter), the ordinary fermion state is
\begin{align}
\begin{aligned}
|\Psi_{0}^{(a)}[g]\rangle
=\mathcal V_{a,M}^{\dagger}[g]
|\Omega_{\mathbf0}^{(M)}\rangle,
\quad
F_i^{\rm can}|\Psi_{0}^{(a)}[g]\rangle
=\delta_{ia}|\Psi_{0}^{(a)}[g]\rangle .
\end{aligned}
\label{eq:ordinary-outgoing-wavepacket-state}
\end{align}
Here $\mathcal V_{a,M}^{\dagger}[g]$ is the vertex operator smeared with a wavepacket $g$:
\begin{align}
\begin{aligned}
\mathcal V_{a,M}^{\dagger}[g]
&=\mathcal N^{-1/2}\int dx\,
g(x)\mathcal V_{a,M}^{\dagger}(x),
\quad\int dx\,|g|^2=1,
\end{aligned}
\label{eq:normalized-smeared-vertex}
\end{align}
with $\mathcal N$ being the normalization factor.
We choose a wavepacket $g$ centered at $X$ with $|X|=R$, energy $E$, and width $\ell\sim E^{-1}$.
Its density is real and even about $X$.

The operator $\mathcal V_{a,M}^{\dagger}(x)$ satisfies the following algebra:
\begin{align}
\begin{aligned}
\bigl[F_i^{\rm can},\mathcal V_{a,M}^{\dagger}(x)\bigr]
&=\delta_{ia}\mathcal V_{a,M}^{\dagger}(x),\\
[b_{i,n},\mathcal V_{a,M}^{\dagger}(x)]
&=\frac{\delta_{ia}}{\sqrt n}e^{-ip_nx}
\mathcal V_{a,M}^{\dagger}(x),\\
[\rho_{i,M}^{\rm can}(y),\mathcal V_{a,M}^{\dagger}(x)]
&=\delta_{ia}\delta_M(y-x)\mathcal V_{a,M}^{\dagger}(x),\\
\delta_M(x)&=\frac{1}{2L}
\left[1+2\sum_{n=1}^{M}\cos(p_nx)\right].
\end{aligned}
\label{eq:finite-current-vertex-algebra}
\end{align}
Note that $\delta_M$ is a regularized delta function,\footnote{The delta function of the circle with the length $2L$ is $\delta_{2L}(x)=\sum_{n=-\infty}^{\infty}e^{ip_nx}/(2L)$.} obtained by keeping the first $M$ Fourier modes.
To construct a semiton, there are two key observations.
First, for $E\ll I^{-1},r_0^{-1}$, Polchinski~\cite{Polchinski:1984uw} showed that the outgoing fermion number is
\begin{align}
\int dt\langle\rho_i^{\rm can}(r_0,t)\rangle
=\delta_{ia}-\frac{1}{2}
\end{align}
for an incoming fermion in the $a$-th flavor, where the time integration is over the first scattering process~\footnote{For finite $L$, there are multiple scattering events, but in this Letter we only focus on the first scattering event.}.
However, the total fermion number in each flavor must remain integral and time independent, which means that the core region must carry the remaining half unit:
\begin{align}
\int_{-r_0}^{r_0} dx\langle\rho_i^{\rm can}(x,t)\rangle=\frac{1}{2}\quad\text{after scattering},
\end{align}
for $i=1,2,3,4$.
Second, Ref.~\cite{Hamada:2022eiv} showed that $\alpha$ shifts by $\pi$ during the scattering.
A detailed derivation is given in Sec.~S3 of the Supplemental Material~\cite{SupplementalMaterial}.

Therefore, to construct a semiton state, we implement a unitary operator that realizes the above two properties.
This is achieved by the following operator:
\begin{align}
\mathcal G^{(M)}[h]
&=e^{-i\pi\Pi}
\exp\!\left[
\sum_{i=1}^{4}\sum_{n=1}^{M}
\frac{h_n^*b_{i,n}^{\dagger}-h_nb_{i,n}}{2\sqrt n}
\right].
\label{eq:joint-rotor-current-displacement}
\end{align}
Here $e^{-i\pi\Pi}$ shifts the rotor coordinate by $\pi$, $h=f-|g|^2$, and $h_{n}=\int_{-L}^{L}dx\,e^{ip_nx}h(x)$.
The physical meaning of the second factor becomes clear from the following algebra:
\begin{align}
\begin{aligned}
\mathcal G^\dagger b_{i,n}\mathcal G
&=b_{i,n}+\frac{h_n^*}{2\sqrt n},
\qquad
\mathcal G^\dagger\alpha\mathcal G=\alpha+\pi,\\
\mathcal G^\dagger\rho_{i,M}^{\rm can}(x)\mathcal G
&=\rho_{i,M}^{\rm can}(x)+\frac{1}{2}h(x),
\qquad
[F_i^{\rm can},\mathcal G]=0,\\
\mathcal G^\dagger J_{i,M}^{\rm cov}(x)\mathcal G
&=J_{i,M}^{\rm cov}(x)-\frac{1}{2}|g(x)|^2.
\end{aligned}
\label{eq:joint-displacement-conjugation}
\end{align}
Namely, the second factor creates a cloud of particle-hole pairs, transferring half a unit per flavor between the wavepacket region and the core without shifting $F_i^{\rm can}$.

\paragraph{Semiton state, energy density, and charges.---}\label{normalized-semiton-ket-density-and-charges}

We define a candidate outgoing semiton state as
\begin{align}
|\Psi_{1/2}^{(a)}[g]\rangle
=\mathcal G^{(M)}[f-|g|^2]
|\Psi_{0}^{(a)}[g]\rangle .
\label{eq:outgoing-semiton-state}
\end{align}
Thus the construction never introduces a half-integral eigenvalue of
$F_i^{\rm can}$.
Eq.~\eqref{eq:joint-displacement-conjugation} gives the identity
\begin{align}
\begin{aligned}
&\langle\Psi_{1/2}^{(a)}[g]|\rho_{i,M}^{\rm can}|\Psi_{1/2}^{(a)}[g]\rangle
-\langle\Psi_{0}^{(a)}[g]|\rho_{i,M}^{\rm can}|\Psi_{0}^{(a)}[g]\rangle
=\frac{1}{2}\left(f-|g|^2\right).
\end{aligned}
\label{eq:finite-cutoff-density-shift}
\end{align}
This indicates that, given the $\rho_{i,M}^{\rm can}(x)$ profile of the ordinary fermion state, we can obtain the $\rho_{i,M}^{\rm can}(x)$ profile of our semiton state.

For the ordinary fermion state, we assume
\begin{align}
&\langle \Psi_{0}^{(a)}[g]|\rho_{i,M}^{\rm can}|\Psi_{0}^{(a)}[g]\rangle
\rightarrow
\delta_{ia}|g|^2,\nonumber\\
&\langle \Psi_{0}^{(a)}[g]|\alpha|\Psi_{0}^{(a)}[g]\rangle
\rightarrow
0.
\label{eq:ordinary-packet-current-density}
\end{align}
The right-hand side of the first line is the expected form of an ordinary fermion.
The second line is consistent with the fact that the rotor is not affected by the wavepacket located far from the core. 
We believe this is true in the limit where the core and wavepacket are infinitely separated while removing the far boundary $x=\pm L$.
Namely, at fixed $r_0$, $I$, $\ell \sim E^{-1}$, and $\Lambda\gg E$, the limit is taken as
\begin{align}
L, R\to\infty,
\qquad \frac{R}{L}\to 0,
\label{eq:ordered-local-limit}
\end{align}
and then $\Lambda\to\infty$.

Combining this with Eq.~\eqref{eq:finite-cutoff-density-shift}, we obtain
\begin{align}
\langle\Psi_{1/2}^{(a)}[g]|\rho_i^{\rm can}(x)|\Psi_{1/2}^{(a)}[g]\rangle
\rightarrow
\frac{1}{2}f(x)
+\left(\delta_{ia}-\frac{1}{2}\right)|g(x)|^2.
\label{eq:semiton-core-front-density}
\end{align}
The first term is common to all four flavors and localized at the core.
The second is precisely the semiton front shown in Table~\ref{tab:chiral-field-charges}.
As expected, $F_i^{\rm can}$ remains integral:
\begin{align}
\int_{-L}^{L}dx\,
\langle\Psi_{1/2}^{(a)}[g]|
\rho_i^{\rm can}(x)
|\Psi_{1/2}^{(a)}[g]\rangle
=\delta_{ia}.
\end{align}

The $U(1)_M$ charge of the whole system is defined as $Q_M:=2\Pi+\int_{-L}^{L}dx\,w_{\rm G}(x)\sum_i\rho_i^{\rm can}(x)$, where we introduce the weight function $w_{\rm G}(x)=1-2\int_{-r_0}^{x}dy\,f(y)$.
It interpolates from $+1$ on the incoming side to $-1$ on the outgoing side.
The core-localized fermion density carries no net $U(1)_M$ charge.
Weighting the core density $f/2$ in each of the four channels gives
\begin{align*}
Q_M^{\rm core}
&:=2\int_{-r_0}^{r_0}dx\,w_{\rm G}(x)f(x)
=-\frac{1}{2}\bigl[w_{\rm G}(x)^2\bigr]_{-r_0}^{r_0}=0.
\end{align*}
On the other hand, the semiton front carries the fermion flavor vector of Table~\ref{tab:chiral-field-charges}.
By using the definition of $Q_M$ and Eq.~\eqref{eq:joint-displacement-conjugation}, we observe that $\mathcal G^{(M)}$ dressing changes the $U(1)_M$ charge of the semiton front by $\pm2$, thereby reproducing $U(1)_M$ weights in the table.
For example, for the outgoing wavepacket, the shift is $+2$, so $Q_M = -1+2 = +1$. 
The dressing does not change the $SU(4)_{\rm f}$ weights.

\begin{figure}
\centering
\includegraphics[width=\columnwidth]{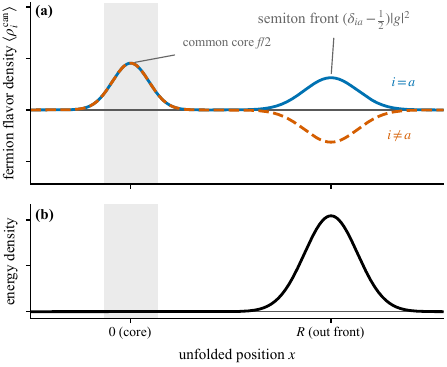}
\caption{Schematic plots of the outgoing semiton fermion number ((a) upper panel) and the energy density ((b) lower panel). (a) The common core term $f/2$ and the semiton front $(\delta_{ia}-\tfrac12)|g|^2$ in channel $i$. (b) The energy density is localized at the outgoing front.}
\label{fig:semiton-profile}
\end{figure}

\paragraph{Energy and asymptotic orthogonality.---}\label{energy-and-infrared-orthogonality}

One may worry that the localized core density in Eq.~\eqref{eq:semiton-core-front-density} costs an energy of order $r_0^{-1}$, i.e., of order the GUT scale.
We show that this is not the case by computing the difference between the expectation values of the Hamiltonian density~\eqref{eq:regulated-current-hamiltonian} in the semiton state and the ordinary fermion state. As $\mathcal G^{(M)}$ commutes with the rotor kinetic term, we consider the remaining part, and define
\begin{align}
\begin{aligned}
\mathcal{H}_{M}(x)&:=\pi\sum_i{:}[J_{i,M}^{\rm cov}(x)]^2{:},\\
\Delta\mathcal E^{(a)}
&:=\langle\Psi_{1/2}^{(a)}[g]|\mathcal{H}_{M}|\Psi_{1/2}^{(a)}[g]\rangle
-\langle\Psi_{0}^{(a)}[g]|\mathcal{H}_{M}|\Psi_{0}^{(a)}[g]\rangle\\
&=-\pi|g|^2
\sum_i\langle\Psi_{0}^{(a)}[g]| J_{i,M}^{\rm cov}|\Psi_{0}^{(a)}[g]\rangle + \pi|g|^4,
\end{aligned}
\label{eq:current-energy-density-definitions}
\end{align}
where Eq.~\eqref{eq:joint-displacement-conjugation} is used.
Applying Eq.~\eqref{eq:ordinary-packet-current-density}, we obtain
\begin{align}
\begin{aligned}
&\Delta\mathcal E^{(a)}
\rightarrow
\pi\left(-1+1\right)|g(x)|^4=0.
\end{aligned}
\label{eq:semiton-energy-density-cancellation}
\end{align}

Eq.~\eqref{eq:semiton-energy-density-cancellation} establishes equality of the energy-density one-point functions and, in particular, the absence of an additional core energy.
The schematic plots of the semiton fermion number and the energy density are summarized in Fig.~\ref{fig:semiton-profile}.

These relations show that the proposed channel is consistent at the level of vacuum-subtracted one-point functions, provided the outgoing packet has the same energy distribution as the incoming one.
Free asymptotic propagation together with Eq.~\eqref{eq:semiton-energy-density-cancellation} then gives equal energy
expectations.
The outgoing semiton carries the same $U(1)_M$ weight $+1$ and the same $SU(4)_{\rm f}$ weights as the ordinary incoming fermion.
The core cloud changes neither result.
Thus the energy and charge expectation values are consistent across the proposed channel.
We remark that this is not a statement about sharp $Q_M$ eigenvalues. 
The vacuum is not a $Q_M$ eigenstate because the boundary condition at $x=L$ breaks the $U(1)_M$ symmetry~\footnote{One way to make the vacuum an eigenstate is to impose an anti-monopole boundary condition, which we leave for future work.}.
Nevertheless, unless the wavepacket goes through the $x=\pm L$ boundary, the expectation values of $Q_M$ are conserved.

To connect this construction with the twisted-sector interpretation, we define the covariant fermion number operator
\begin{align}
F_i^{\rm cov}
:=\int_{-L}^{L}dx\,J_{i,M}^{\rm cov}(x)
=F_i^{\rm can}-\frac{\alpha}{2\pi}.
\label{eq:covariant-fermion-number}
\end{align}
From Eq.~\eqref{eq:joint-displacement-conjugation}, in terms of $F_i^{\rm cov}$, our semiton states have $\langle F_i^{\rm cov}\rangle=\delta_{ia}-1/2$, corresponding to the outgoing semiton flavor vector in Table~\ref{tab:chiral-field-charges}.
This matches the picture in the literature, although the low-energy state is far from an eigenstate of $F_i^{\rm cov}$ as $\alpha$ fluctuates in the vacuum.

We can generalize our construction to general $N_f$ flavors by replacing the operator $\mathcal G^{(M)}$ with
\begin{align*}
\mathcal G_{N_f}^{(M)}=
e^{-i\frac{4\pi}{N_f}\Pi}
\exp\!\left[
\frac{2}{N_f}
\sum_{i=1}^{N_f}\sum_{n=1}^{M}
\frac{h_n^*b_{i,n}^{\dagger}-h_nb_{i,n}}{\sqrt n}
\right].
\end{align*}
For example, for $N_f=2$, the operation involves a $2\pi$ rotor shift, implying that the resulting state is an excitation of the vacuum in a different spectral-flow branch, as discussed in Ref.~\cite{Hamada:2022eiv}.

Finally, we consider $\langle\Psi_0^{(a)}[g]|\Psi_{1/2}^{(a)}[g]\rangle$, the overlap of the ordinary and semiton states constructed from the same wavepacket.
Under the point-excitation approximation stated in Sec.~S2 of the
Supplemental Material~\cite{SupplementalMaterial}, we obtain the bound
\begin{align}
\left|\langle\Psi_0^{(a)}[g]|\Psi_{1/2}^{(a)}[g]\rangle\right|\leq
\text{const.}\times\left(\Lambda R\right)^{-1/4}
\rightarrow0,
\label{eq:anderson-overlap-scaling}
\end{align}
for $R\to\infty$.
The orthogonality comes from the IR effect~\cite{Anderson:1967zze}.

\paragraph{Conclusion.---}
We have constructed an explicit state for the semiton in the fermion--rotor model.
The vacuum is a state that entangles the rotor with the Dirac sea.
We have constructed an operator that transforms an ordinary fermion state into a semiton state with a core cloud.
This particle--hole dressing may be compatible with soliton descriptions~\cite{Kitano:2021pwt} of the semiton state based on multifermion condensates~\cite{Kazama:1983xp}.
We do not derive the constructed semiton state from the real-time
evolution of the incoming state, but we show that the outgoing semiton state has the same energy and charge expectation values as the ordinary incoming fermion state.
Moreover, the ordinary and semiton states are asymptotically orthogonal within the point-excitation approximation.

\begin{acknowledgments}
This work was supported in part by JSPS KAKENHI Grant Nos. JP24H00976, JP24K07035, and JP24KF0167, and by JST BOOST Program Japan Grant No. JPMJBY25E1.

\textit{AI use statement.---} OpenAI Codex (GPT-5.6-sol) was used for computations in the Letter. In particular, the tool initially suggested the form of $\mathcal G^{(M)}$ in Eq.~\eqref{eq:joint-rotor-current-displacement}.  The author independently verified its unitarity, the conjugation identities, and all conclusions drawn from them.  The tool also generated the plotting code for the schematic plots in Fig.~\ref{fig:semiton-profile} from instructions supplied by the author.  The author is fully responsible for all scientific claims and the manuscript.
\end{acknowledgments}

\bibliography{semiton_state}

\begin{thebibliography}{33}%
\makeatletter
\providecommand \@ifxundefined [1]{%
 \@ifx{#1\undefined}
}%
\providecommand \@ifnum [1]{%
 \ifnum #1\expandafter \@firstoftwo
 \else \expandafter \@secondoftwo
 \fi
}%
\providecommand \@ifx [1]{%
 \ifx #1\expandafter \@firstoftwo
 \else \expandafter \@secondoftwo
 \fi
}%
\providecommand \natexlab [1]{#1}%
\providecommand \enquote  [1]{``#1''}%
\providecommand \bibnamefont  [1]{#1}%
\providecommand \bibfnamefont [1]{#1}%
\providecommand \citenamefont [1]{#1}%
\providecommand \href@noop [0]{\@secondoftwo}%
\providecommand \href [0]{\begingroup \@sanitize@url \@href}%
\providecommand \@href[1]{\@@startlink{#1}\@@href}%
\providecommand \@@href[1]{\endgroup#1\@@endlink}%
\providecommand \@sanitize@url [0]{\catcode `\\12\catcode `\$12\catcode
  `\&12\catcode `\#12\catcode `\^12\catcode `\_12\catcode `\%12\relax}%
\providecommand \@@startlink[1]{}%
\providecommand \@@endlink[0]{}%
\providecommand \url  [0]{\begingroup\@sanitize@url \@url }%
\providecommand \@url [1]{\endgroup\@href {#1}{\urlprefix }}%
\providecommand \urlprefix  [0]{URL }%
\providecommand \Eprint [0]{\href }%
\providecommand \doibase [0]{https://doi.org/}%
\providecommand \selectlanguage [0]{\@gobble}%
\providecommand \bibinfo  [0]{\@secondoftwo}%
\providecommand \bibfield  [0]{\@secondoftwo}%
\providecommand \translation [1]{[#1]}%
\providecommand \BibitemOpen [0]{}%
\providecommand \bibitemStop [0]{}%
\providecommand \bibitemNoStop [0]{.\EOS\space}%
\providecommand \EOS [0]{\spacefactor3000\relax}%
\providecommand \BibitemShut  [1]{\csname bibitem#1\endcsname}%
\let\auto@bib@innerbib\@empty
\bibitem [{\citenamefont {'t~Hooft}(1974)}]{tHooft:1974kcl}%
  \BibitemOpen
  \bibfield  {author} {\bibinfo {author} {\bibfnamefont {G.}~\bibnamefont
  {'t~Hooft}},\ }\bibfield  {title} {\bibinfo {title} {{Magnetic Monopoles in
  Unified Gauge Theories}},\ }\href
  {https://doi.org/10.1016/0550-3213(74)90486-6} {\bibfield  {journal}
  {\bibinfo  {journal} {Nucl. Phys. B}\ }\textbf {\bibinfo {volume} {79}},\
  \bibinfo {pages} {276} (\bibinfo {year} {1974})}\BibitemShut {NoStop}%
\bibitem [{\citenamefont {Polyakov}(1974)}]{Polyakov:1974ek}%
  \BibitemOpen
  \bibfield  {author} {\bibinfo {author} {\bibfnamefont {A.~M.}\ \bibnamefont
  {Polyakov}},\ }\bibfield  {title} {\bibinfo {title} {{Particle Spectrum in
  Quantum Field Theory}},\ }\href@noop {} {\bibfield  {journal} {\bibinfo
  {journal} {JETP Lett.}\ }\textbf {\bibinfo {volume} {20}},\ \bibinfo {pages}
  {194} (\bibinfo {year} {1974})}\BibitemShut {NoStop}%
\bibitem [{\citenamefont {Rubakov}(1982)}]{Rubakov:1982fp}%
  \BibitemOpen
  \bibfield  {author} {\bibinfo {author} {\bibfnamefont {V.~A.}\ \bibnamefont
  {Rubakov}},\ }\bibfield  {title} {\bibinfo {title} {{Adler-Bell-Jackiw
  Anomaly and Fermion Number Breaking in the Presence of a Magnetic
  Monopole}},\ }\href {https://doi.org/10.1016/0550-3213(82)90034-7} {\bibfield
   {journal} {\bibinfo  {journal} {Nucl. Phys. B}\ }\textbf {\bibinfo {volume}
  {203}},\ \bibinfo {pages} {311} (\bibinfo {year} {1982})}\BibitemShut
  {NoStop}%
\bibitem [{\citenamefont {Callan}(1982{\natexlab{a}})}]{Callan:1982au}%
  \BibitemOpen
  \bibfield  {author} {\bibinfo {author} {\bibfnamefont {C.~G.}\ \bibnamefont
  {Callan}, \bibfnamefont {Jr.}},\ }\bibfield  {title} {\bibinfo {title}
  {{Dyon-Fermion Dynamics}},\ }\href {https://doi.org/10.1103/PhysRevD.26.2058}
  {\bibfield  {journal} {\bibinfo  {journal} {Phys. Rev. D}\ }\textbf {\bibinfo
  {volume} {26}},\ \bibinfo {pages} {2058} (\bibinfo {year}
  {1982}{\natexlab{a}})}\BibitemShut {NoStop}%
\bibitem [{\citenamefont {Callan}(1982{\natexlab{b}})}]{Callan:1982ah}%
  \BibitemOpen
  \bibfield  {author} {\bibinfo {author} {\bibfnamefont {C.~G.}\ \bibnamefont
  {Callan}, \bibfnamefont {Jr.}},\ }\bibfield  {title} {\bibinfo {title}
  {{Disappearing Dyons}},\ }\href {https://doi.org/10.1103/PhysRevD.25.2141}
  {\bibfield  {journal} {\bibinfo  {journal} {Phys. Rev. D}\ }\textbf {\bibinfo
  {volume} {25}},\ \bibinfo {pages} {2141} (\bibinfo {year}
  {1982}{\natexlab{b}})}\BibitemShut {NoStop}%
\bibitem [{\citenamefont {Callan}(1983{\natexlab{a}})}]{Callan:1982ac}%
  \BibitemOpen
  \bibfield  {author} {\bibinfo {author} {\bibfnamefont {C.~G.}\ \bibnamefont
  {Callan}, \bibfnamefont {Jr.}},\ }\bibfield  {title} {\bibinfo {title}
  {{Monopole Catalysis of Baryon Decay}},\ }\href
  {https://doi.org/10.1016/0550-3213(83)90677-6} {\bibfield  {journal}
  {\bibinfo  {journal} {Nucl. Phys. B}\ }\textbf {\bibinfo {volume} {212}},\
  \bibinfo {pages} {391} (\bibinfo {year} {1983}{\natexlab{a}})}\BibitemShut
  {NoStop}%
\bibitem [{\citenamefont {Callan}(1983{\natexlab{b}})}]{Callan:1983tm}%
  \BibitemOpen
  \bibfield  {author} {\bibinfo {author} {\bibfnamefont {C.~g.}\ \bibnamefont
  {Callan}, \bibfnamefont {Jr.}},\ }\bibfield  {title} {\bibinfo {title} {{THE
  MONOPOLE CATALYSIS S MATRIX}},\ }in\ \href {https://doi.org/10.1063/1.34591}
  {\emph {\bibinfo {booktitle} {{Workshop on Problems in Unification and
  Supergravity}}}}\ (\bibinfo {year} {1983})\ pp.\ \bibinfo {pages}
  {45--53}\BibitemShut {NoStop}%
\bibitem [{\citenamefont {Polchinski}(1984)}]{Polchinski:1984uw}%
  \BibitemOpen
  \bibfield  {author} {\bibinfo {author} {\bibfnamefont {J.}~\bibnamefont
  {Polchinski}},\ }\bibfield  {title} {\bibinfo {title} {{Monopole Catalysis:
  The Fermion Rotor System}},\ }\href
  {https://doi.org/10.1016/0550-3213(84)90398-5} {\bibfield  {journal}
  {\bibinfo  {journal} {Nucl. Phys. B}\ }\textbf {\bibinfo {volume} {242}},\
  \bibinfo {pages} {345} (\bibinfo {year} {1984})}\BibitemShut {NoStop}%
\bibitem [{\citenamefont {Affleck}\ and\ \citenamefont
  {Sagi}(1994)}]{AffleckSagi:1993bcft}%
  \BibitemOpen
  \bibfield  {author} {\bibinfo {author} {\bibfnamefont {I.}~\bibnamefont
  {Affleck}}\ and\ \bibinfo {author} {\bibfnamefont {J.}~\bibnamefont {Sagi}},\
  }\bibfield  {title} {\bibinfo {title} {{Monopole Catalysed Baryon Decay: A
  Boundary Conformal Field Theory Approach}},\ }\href
  {https://doi.org/10.1016/0550-3213(94)90478-2} {\bibfield  {journal}
  {\bibinfo  {journal} {Nucl. Phys. B}\ }\textbf {\bibinfo {volume} {417}},\
  \bibinfo {pages} {374} (\bibinfo {year} {1994})},\ \Eprint
  {https://arxiv.org/abs/hep-th/9311056} {arXiv:hep-th/9311056} \BibitemShut
  {NoStop}%
\bibitem [{\citenamefont {Yegulalp}(1994)}]{Yegulalp:1994eq}%
  \BibitemOpen
  \bibfield  {author} {\bibinfo {author} {\bibfnamefont {A.}~\bibnamefont
  {Yegulalp}},\ }\bibfield  {title} {\bibinfo {title} {{Fermions Coupled to a
  Conformal Boundary: A Generalization of the Monopole-Fermion System}},\
  }\href {https://doi.org/10.1016/0370-2693(94)91494-X} {\bibfield  {journal}
  {\bibinfo  {journal} {Phys. Lett. B}\ }\textbf {\bibinfo {volume} {328}},\
  \bibinfo {pages} {379} (\bibinfo {year} {1994})},\ \Eprint
  {https://arxiv.org/abs/hep-th/9403125} {arXiv:hep-th/9403125} \BibitemShut
  {NoStop}%
\bibitem [{\citenamefont {Maldacena}\ and\ \citenamefont
  {Ludwig}(1997)}]{Maldacena:1995pq}%
  \BibitemOpen
  \bibfield  {author} {\bibinfo {author} {\bibfnamefont {J.~M.}\ \bibnamefont
  {Maldacena}}\ and\ \bibinfo {author} {\bibfnamefont {A.~W.~W.}\ \bibnamefont
  {Ludwig}},\ }\bibfield  {title} {\bibinfo {title} {{Majorana fermions, exact
  mapping between quantum impurity fixed points with four bulk fermion species,
  and solution of the 'unitarity puzzle'}},\ }\href
  {https://doi.org/10.1016/S0550-3213(97)00596-8} {\bibfield  {journal}
  {\bibinfo  {journal} {Nucl. Phys. B}\ }\textbf {\bibinfo {volume} {506}},\
  \bibinfo {pages} {565} (\bibinfo {year} {1997})},\ \Eprint
  {https://arxiv.org/abs/cond-mat/9502109} {arXiv:cond-mat/9502109}
  \BibitemShut {NoStop}%
\bibitem [{\citenamefont {Kitano}\ and\ \citenamefont
  {Matsudo}(2022)}]{Kitano:2021pwt}%
  \BibitemOpen
  \bibfield  {author} {\bibinfo {author} {\bibfnamefont {R.}~\bibnamefont
  {Kitano}}\ and\ \bibinfo {author} {\bibfnamefont {R.}~\bibnamefont
  {Matsudo}},\ }\bibfield  {title} {\bibinfo {title} {{Missing final state
  puzzle in the monopole-fermion scattering}},\ }\href
  {https://doi.org/10.1016/j.physletb.2022.137271} {\bibfield  {journal}
  {\bibinfo  {journal} {Phys. Lett. B}\ }\textbf {\bibinfo {volume} {832}},\
  \bibinfo {pages} {137271} (\bibinfo {year} {2022})},\ \Eprint
  {https://arxiv.org/abs/2103.13639} {arXiv:2103.13639 [hep-th]} \BibitemShut
  {NoStop}%
\bibitem [{\citenamefont {Csaki}\ \emph {et~al.}(2022)\citenamefont {Csaki},
  \citenamefont {Shirman}, \citenamefont {Telem},\ and\ \citenamefont
  {Terning}}]{Csaki:2021ozp}%
  \BibitemOpen
  \bibfield  {author} {\bibinfo {author} {\bibfnamefont {C.}~\bibnamefont
  {Csaki}}, \bibinfo {author} {\bibfnamefont {Y.}~\bibnamefont {Shirman}},
  \bibinfo {author} {\bibfnamefont {O.}~\bibnamefont {Telem}},\ and\ \bibinfo
  {author} {\bibfnamefont {J.}~\bibnamefont {Terning}},\ }\bibfield  {title}
  {\bibinfo {title} {{Pairwise Multiparticle States and the Monopole Unitarity
  Puzzle}},\ }\href {https://doi.org/10.1103/PhysRevLett.129.181601} {\bibfield
   {journal} {\bibinfo  {journal} {Phys. Rev. Lett.}\ }\textbf {\bibinfo
  {volume} {129}},\ \bibinfo {pages} {181601} (\bibinfo {year} {2022})},\
  \Eprint {https://arxiv.org/abs/2109.01145} {arXiv:2109.01145 [hep-th]}
  \BibitemShut {NoStop}%
\bibitem [{\citenamefont {Brennan}(2023)}]{Brennan:2021ewu}%
  \BibitemOpen
  \bibfield  {author} {\bibinfo {author} {\bibfnamefont {T.~D.}\ \bibnamefont
  {Brennan}},\ }\bibfield  {title} {\bibinfo {title} {{Callan-Rubakov effect
  and higher charge monopoles}},\ }\href
  {https://doi.org/10.1007/JHEP02(2023)159} {\bibfield  {journal} {\bibinfo
  {journal} {JHEP}\ }\textbf {\bibinfo {volume} {02}},\ \bibinfo {pages}
  {159}},\ \Eprint {https://arxiv.org/abs/2109.11207} {arXiv:2109.11207
  [hep-th]} \BibitemShut {NoStop}%
\bibitem [{\citenamefont {Hamada}\ \emph {et~al.}(2022)\citenamefont {Hamada},
  \citenamefont {Kitahara},\ and\ \citenamefont {Sato}}]{Hamada:2022eiv}%
  \BibitemOpen
  \bibfield  {author} {\bibinfo {author} {\bibfnamefont {Y.}~\bibnamefont
  {Hamada}}, \bibinfo {author} {\bibfnamefont {T.}~\bibnamefont {Kitahara}},\
  and\ \bibinfo {author} {\bibfnamefont {Y.}~\bibnamefont {Sato}},\ }\bibfield
  {title} {\bibinfo {title} {{Monopole-fermion scattering and varying Fock
  space}},\ }\href {https://doi.org/10.1007/JHEP11(2022)116} {\bibfield
  {journal} {\bibinfo  {journal} {JHEP}\ }\textbf {\bibinfo {volume} {11}},\
  \bibinfo {pages} {116}},\ \Eprint {https://arxiv.org/abs/2208.01052}
  {arXiv:2208.01052 [hep-th]} \BibitemShut {NoStop}%
\bibitem [{\citenamefont {van Beest}\ \emph {et~al.}(2025)\citenamefont {van
  Beest}, \citenamefont {Boyle~Smith}, \citenamefont {Delmastro}, \citenamefont
  {Komargodski},\ and\ \citenamefont {Tong}}]{vanBeest:2023dbu}%
  \BibitemOpen
  \bibfield  {author} {\bibinfo {author} {\bibfnamefont {M.}~\bibnamefont {van
  Beest}}, \bibinfo {author} {\bibfnamefont {P.}~\bibnamefont {Boyle~Smith}},
  \bibinfo {author} {\bibfnamefont {D.}~\bibnamefont {Delmastro}}, \bibinfo
  {author} {\bibfnamefont {Z.}~\bibnamefont {Komargodski}},\ and\ \bibinfo
  {author} {\bibfnamefont {D.}~\bibnamefont {Tong}},\ }\bibfield  {title}
  {\bibinfo {title} {{Monopoles, scattering, and generalized symmetries}},\
  }\href {https://doi.org/10.1007/JHEP03(2025)014} {\bibfield  {journal}
  {\bibinfo  {journal} {JHEP}\ }\textbf {\bibinfo {volume} {03}},\ \bibinfo
  {pages} {014}},\ \Eprint {https://arxiv.org/abs/2306.07318} {arXiv:2306.07318
  [hep-th]} \BibitemShut {NoStop}%
\bibitem [{\citenamefont {Khoze}(2023)}]{Khoze:2023kiu}%
  \BibitemOpen
  \bibfield  {author} {\bibinfo {author} {\bibfnamefont {V.~V.}\ \bibnamefont
  {Khoze}},\ }\bibfield  {title} {\bibinfo {title} {{Scattering amplitudes of
  fermions on monopoles}},\ }\href {https://doi.org/10.1007/JHEP11(2023)214}
  {\bibfield  {journal} {\bibinfo  {journal} {JHEP}\ }\textbf {\bibinfo
  {volume} {11}},\ \bibinfo {pages} {214}},\ \Eprint
  {https://arxiv.org/abs/2308.09401} {arXiv:2308.09401 [hep-th]} \BibitemShut
  {NoStop}%
\bibitem [{\citenamefont {Brennan}(2024)}]{Brennan:2023tae}%
  \BibitemOpen
  \bibfield  {author} {\bibinfo {author} {\bibfnamefont {T.~D.}\ \bibnamefont
  {Brennan}},\ }\bibfield  {title} {\bibinfo {title} {{A new solution to the
  Callan Rubakov effect}},\ }\href {https://doi.org/10.1007/JHEP11(2024)170}
  {\bibfield  {journal} {\bibinfo  {journal} {JHEP}\ }\textbf {\bibinfo
  {volume} {11}},\ \bibinfo {pages} {170}},\ \Eprint
  {https://arxiv.org/abs/2309.00680} {arXiv:2309.00680 [hep-th]} \BibitemShut
  {NoStop}%
\bibitem [{\citenamefont {van Beest}\ \emph {et~al.}(2024)\citenamefont {van
  Beest}, \citenamefont {Boyle~Smith}, \citenamefont {Delmastro}, \citenamefont
  {Mouland},\ and\ \citenamefont {Tong}}]{vanBeest:2023mbs}%
  \BibitemOpen
  \bibfield  {author} {\bibinfo {author} {\bibfnamefont {M.}~\bibnamefont {van
  Beest}}, \bibinfo {author} {\bibfnamefont {P.}~\bibnamefont {Boyle~Smith}},
  \bibinfo {author} {\bibfnamefont {D.}~\bibnamefont {Delmastro}}, \bibinfo
  {author} {\bibfnamefont {R.}~\bibnamefont {Mouland}},\ and\ \bibinfo {author}
  {\bibfnamefont {D.}~\bibnamefont {Tong}},\ }\bibfield  {title} {\bibinfo
  {title} {{Fermion-monopole scattering in the Standard Model}},\ }\href
  {https://doi.org/10.1007/JHEP08(2024)004} {\bibfield  {journal} {\bibinfo
  {journal} {JHEP}\ }\textbf {\bibinfo {volume} {08}},\ \bibinfo {pages}
  {004}},\ \Eprint {https://arxiv.org/abs/2312.17746} {arXiv:2312.17746
  [hep-th]} \BibitemShut {NoStop}%
\bibitem [{\citenamefont {Loladze}\ and\ \citenamefont
  {Okui}(2025)}]{Loladze:2024ayk}%
  \BibitemOpen
  \bibfield  {author} {\bibinfo {author} {\bibfnamefont {V.}~\bibnamefont
  {Loladze}}\ and\ \bibinfo {author} {\bibfnamefont {T.}~\bibnamefont {Okui}},\
  }\bibfield  {title} {\bibinfo {title} {{Monopole-Fermion Scattering and the
  Solution to the Semiton{\textendash}Unitarity Puzzle}},\ }\href
  {https://doi.org/10.1103/PhysRevLett.134.051602} {\bibfield  {journal}
  {\bibinfo  {journal} {Phys. Rev. Lett.}\ }\textbf {\bibinfo {volume} {134}},\
  \bibinfo {pages} {051602} (\bibinfo {year} {2025})},\ \Eprint
  {https://arxiv.org/abs/2408.04577} {arXiv:2408.04577 [hep-th]} \BibitemShut
  {NoStop}%
\bibitem [{\citenamefont {Loladze}\ \emph {et~al.}(2026)\citenamefont
  {Loladze}, \citenamefont {Okui},\ and\ \citenamefont
  {Tong}}]{Loladze:2025jsq}%
  \BibitemOpen
  \bibfield  {author} {\bibinfo {author} {\bibfnamefont {V.}~\bibnamefont
  {Loladze}}, \bibinfo {author} {\bibfnamefont {T.}~\bibnamefont {Okui}},\ and\
  \bibinfo {author} {\bibfnamefont {D.}~\bibnamefont {Tong}},\ }\bibfield
  {title} {\bibinfo {title} {{Dynamics of the fermion-rotor system}},\ }\href
  {https://doi.org/10.1007/JHEP01(2026)052} {\bibfield  {journal} {\bibinfo
  {journal} {JHEP}\ }\textbf {\bibinfo {volume} {01}},\ \bibinfo {pages}
  {052}},\ \Eprint {https://arxiv.org/abs/2508.21059} {arXiv:2508.21059
  [hep-th]} \BibitemShut {NoStop}%
\bibitem [{\citenamefont {Tachikawa}\ \emph {et~al.}(2026)\citenamefont
  {Tachikawa}, \citenamefont {Tsuji},\ and\ \citenamefont
  {Watanabe}}]{Tachikawa:2026cxd}%
  \BibitemOpen
  \bibfield  {author} {\bibinfo {author} {\bibfnamefont {Y.}~\bibnamefont
  {Tachikawa}}, \bibinfo {author} {\bibfnamefont {K.}~\bibnamefont {Tsuji}},\
  and\ \bibinfo {author} {\bibfnamefont {M.}~\bibnamefont {Watanabe}},\
  }\href@noop {} {\bibinfo {title} {{What happens to wavepackets of fermions
  when scattered by the Maldacena-Ludwig wall?}}} (\bibinfo {year} {2026}),\
  \bibinfo {note} {{arXiv:2603.25508}},\ \Eprint
  {https://arxiv.org/abs/2603.25508} {arXiv:2603.25508 [hep-th]} \BibitemShut
  {NoStop}%
\bibitem [{\citenamefont {Smith}\ and\ \citenamefont
  {Tong}(2020)}]{Smith:2020nuf}%
  \BibitemOpen
  \bibfield  {author} {\bibinfo {author} {\bibfnamefont {P.~B.}\ \bibnamefont
  {Smith}}\ and\ \bibinfo {author} {\bibfnamefont {D.}~\bibnamefont {Tong}},\
  }\href@noop {} {\bibinfo {title} {{What Symmetries are Preserved by a Fermion
  Boundary State?}}} (\bibinfo {year} {2020}),\ \bibinfo {note}
  {{arXiv:2006.07369}},\ \Eprint {https://arxiv.org/abs/2006.07369}
  {arXiv:2006.07369 [hep-th]} \BibitemShut {NoStop}%
\bibitem [{\citenamefont {Kazama}\ \emph {et~al.}(1977)\citenamefont {Kazama},
  \citenamefont {Yang},\ and\ \citenamefont {Goldhaber}}]{Kazama:1976fm}%
  \BibitemOpen
  \bibfield  {author} {\bibinfo {author} {\bibfnamefont {Y.}~\bibnamefont
  {Kazama}}, \bibinfo {author} {\bibfnamefont {C.~N.}\ \bibnamefont {Yang}},\
  and\ \bibinfo {author} {\bibfnamefont {A.~S.}\ \bibnamefont {Goldhaber}},\
  }\bibfield  {title} {\bibinfo {title} {{Scattering of a Dirac Particle with
  Charge Ze by a Fixed Magnetic Monopole}},\ }\href
  {https://doi.org/10.1103/PhysRevD.15.2287} {\bibfield  {journal} {\bibinfo
  {journal} {Phys. Rev. D}\ }\textbf {\bibinfo {volume} {15}},\ \bibinfo
  {pages} {2287} (\bibinfo {year} {1977})}\BibitemShut {NoStop}%
\bibitem [{Note1()}]{Note1}%
  \BibitemOpen
  \bibinfo {note} {To be precise, this boundary condition breaks the $U(1)_M$
  symmetry introduced below. We will discuss this point later.}\BibitemShut
  {Stop}%
\bibitem [{Sup()}]{SupplementalMaterial}%
  \BibitemOpen
  \href@noop {} {}\bibinfo {note} {See Supplemental Material for the derivation
  of the current Hamiltonian from the bosonized boundary action, the exact
  finite-regulator overlap identity and the conditional asymptotic overlap
  bound in the point-excitation approximation, and the retarded Ward derivation
  of the half-unit core density}\BibitemShut {NoStop}%
\bibitem [{Note2()}]{Note2}%
  \BibitemOpen
  \bibinfo {note} {The constant term arises from the commutation relation
  between $x_n$ and $\pi _n$.}\BibitemShut {Stop}%
\bibitem [{\citenamefont {von Delft}\ and\ \citenamefont
  {Schoeller}(1998)}]{vonDelftSchoeller:1998bosonization}%
  \BibitemOpen
  \bibfield  {author} {\bibinfo {author} {\bibfnamefont {J.}~\bibnamefont {von
  Delft}}\ and\ \bibinfo {author} {\bibfnamefont {H.}~\bibnamefont
  {Schoeller}},\ }\bibfield  {title} {\bibinfo {title} {{Bosonization for
  Beginners---Refermionization for Experts}},\ }\href
  {https://doi.org/10.1002/andp.19985100401} {\bibfield  {journal} {\bibinfo
  {journal} {Annalen Phys.}\ }\textbf {\bibinfo {volume} {7}},\ \bibinfo
  {pages} {225} (\bibinfo {year} {1998})},\ \Eprint
  {https://arxiv.org/abs/cond-mat/9805275} {arXiv:cond-mat/9805275}
  \BibitemShut {NoStop}%
\bibitem [{Note3()}]{Note3}%
  \BibitemOpen
  \bibinfo {note} {The delta function of the circle with the length $2L$ is
  $\delta _{2L}(x)=\DOTSB \sum@ \slimits@ _{n=-\infty }^{\infty
  }e^{ip_nx}/(2L)$.}\BibitemShut {Stop}%
\bibitem [{Note4()}]{Note4}%
  \BibitemOpen
  \bibinfo {note} {For finite $L$, there are multiple scattering events, but in
  this Letter we only focus on the first scattering event.}\BibitemShut {Stop}%
\bibitem [{Note5()}]{Note5}%
  \BibitemOpen
  \bibinfo {note} {One way to make the vacuum an eigenstate is to impose an
  anti-monopole boundary condition, which we leave for future
  work.}\BibitemShut {Stop}%
\bibitem [{\citenamefont {Anderson}(1967)}]{Anderson:1967zze}%
  \BibitemOpen
  \bibfield  {author} {\bibinfo {author} {\bibfnamefont {P.~W.}\ \bibnamefont
  {Anderson}},\ }\bibfield  {title} {\bibinfo {title} {{Infrared Catastrophe in
  Fermi Gases with Local Scattering Potentials}},\ }\href
  {https://doi.org/10.1103/PhysRevLett.18.1049} {\bibfield  {journal} {\bibinfo
   {journal} {Phys. Rev. Lett.}\ }\textbf {\bibinfo {volume} {18}},\ \bibinfo
  {pages} {1049} (\bibinfo {year} {1967})}\BibitemShut {NoStop}%
\bibitem [{\citenamefont {Kazama}(1983)}]{Kazama:1983xp}%
  \BibitemOpen
  \bibfield  {author} {\bibinfo {author} {\bibfnamefont {Y.}~\bibnamefont
  {Kazama}},\ }\bibfield  {title} {\bibinfo {title} {{Condensates and the
  Boundary Condition in Monopole - Fermion Dynamics}},\ }\href
  {https://doi.org/10.1143/PTP.70.1166} {\bibfield  {journal} {\bibinfo
  {journal} {Prog. Theor. Phys.}\ }\textbf {\bibinfo {volume} {70}},\ \bibinfo
  {pages} {1166} (\bibinfo {year} {1983})}\BibitemShut {NoStop}%
\end{thebibliography}%

\section*{End Matter}

\subsection{Spectral Flow, Theta Vacuum and Semiton States}\label{sec:endmatter-global-compact-states}
The Hamiltonian~\eqref{eq:regulated-current-hamiltonian} satisfies the following spectral-flow gluing condition:
\begin{align}
\begin{aligned}
H_{\rm cur}^{(M)}(\alpha+2\pi)
&=\widehat{\mathcal U}_{2\pi}^{(M)}
H_{\rm cur}^{(M)}(\alpha)
\bigl(\widehat{\mathcal U}_{2\pi}^{(M)}\bigr)^{-1},\\
\widehat{\mathcal U}_{2\pi}^{(M)}
&=S_{\mathbf q}\prod_{i=1}^{4}\prod_{n=1}^{M}
\exp\!\left[\frac{f_n}{\sqrt n}
(b_{i,n}^{\dagger}-b_{i,n})\right].
\end{aligned}
\label{eq:endmatter-spectral-flow-unitary}
\end{align}
where $S_{\mathbf q}|\mathbf n\rangle=|\mathbf n+\mathbf q\rangle$, and ${\mathbf q}=(1,1,1,1)$.
The operator also obeys
\begin{align}
\begin{aligned}
(\widehat{\mathcal U}_{2\pi}^{(M)})^{-1}F_i^{\rm can}
\widehat{\mathcal U}_{2\pi}^{(M)}=F_i^{\rm can}+1,
\\
(\widehat{\mathcal U}_{2\pi}^{(M)})^{-1}b_{i,n}
\widehat{\mathcal U}_{2\pi}^{(M)}
=b_{i,n}+\frac{f_n}{\sqrt n}.
\end{aligned}
\label{eq:endmatter-spectral-flow-action}
\end{align}

The covariant fermion-number charge $F_i^{\rm cov}(\alpha)$, defined in Eq.~\eqref{eq:covariant-fermion-number}, also has a spectral-flow connection:
\begin{align}
\begin{aligned}
\bigl(\widehat{\mathcal U}_{2\pi}^{(M)}\bigr)^{-1}
F_i^{\rm cov}(\alpha+2\pi)
\widehat{\mathcal U}_{2\pi}^{(M)}
&=F_i^{\rm cov}(\alpha).
\end{aligned}
\label{eq:endmatter-covariant-charge-gluing}
\end{align}
We thus see that there are two distinct fermion flavor charges, $F_i^{\rm can}$ and $F_i^{\rm cov}$.
The former is always integer-valued and well-defined on one branch, but changes by one unit under spectral flow.
The latter is globally well-defined, but low-energy states are not its eigenstates, because $\alpha$ fluctuates.

We now use Eqs.~\eqref{eq:endmatter-spectral-flow-unitary}--\eqref{eq:endmatter-covariant-charge-gluing} to construct the global compact states explicitly, choosing the $\theta=0$ sector.
We denote the unit vector in flavor $a$ by $\mathbf e_a$.
The lattice of $\mathbf F^{\rm can}$ decomposes into spectral-flow orbits,
and we define
\begin{align}
\begin{aligned}
\mathcal O_{\mathbf r}
&:=\{\mathbf r+k\mathbf q:k\in\mathbb Z\},
\\
\boldsymbol\Phi_{\mathbf0}^{(M)}(\alpha)
&:={}_{\rm rot}\langle\alpha|\Omega_{\mathbf0}^{(M)}\rangle\in\mathcal H_{\rm cur}^{(M)},
\end{aligned}
\label{eq:endmatter-spectral-flow-orbit-seed}
\end{align}
where $|\Omega_{\mathbf0}^{(M)}\rangle$ is defined in Eq.~\eqref{eq:branch-local-vacuum-ket}, and the vacuum orbit is $\mathcal O_{\mathbf0}$.
Then, the theta vacuum is constructed by summing over the orbit:
\begin{align}
\begin{aligned}
\boldsymbol\Xi_{\mathbf0}^{(M)}(\alpha)
&:=\sum_{k\in\mathbb Z}
\bigl(\widehat{\mathcal U}_{2\pi}^{(M)}\bigr)^k
\boldsymbol\Phi_{\mathbf0}^{(M)}(\alpha-2\pi k),
\\
|\Omega_{\theta=0}^{(M)}\rangle
&:=\int_{-\pi}^{\pi}d\alpha\,
|\alpha\rangle_{\rm rot}\otimes
\boldsymbol\Xi_{\mathbf0}^{(M)}(\alpha).
\end{aligned}
\label{eq:endmatter-global-theta-vacuum}
\end{align}
Its $k$-th summand lies in the branch $k\mathbf q$.
Since $\widehat{\mathcal U}_{2\pi}^{(M)}$ is unitary and states with different eigenvalues of $\mathbf F^{\rm can}$ are orthogonal, the state is normalized:
\begin{align}
\begin{aligned}
\int_{-\pi}^{\pi}d\alpha\,
\|\boldsymbol\Xi_{\mathbf0}^{(M)}(\alpha)\|^2
&=\sum_{k\in\mathbb Z}\int_{-\pi}^{\pi}d\alpha\,
\|\boldsymbol\Phi_{\mathbf0}^{(M)}(\alpha-2\pi k)\|^2\\
&=\int_{\mathbb R}d\widetilde\alpha\,
\|\boldsymbol\Phi_{\mathbf0}^{(M)}(\widetilde\alpha)\|^2=1.
\end{aligned}
\label{eq:endmatter-global-vacuum-normalization}
\end{align}
One can show that $\boldsymbol\Xi_{\mathbf0}^{(M)}$ satisfies
\begin{align}
\begin{aligned}
\boldsymbol\Xi_{\mathbf0}^{(M)}(\pi)
&=\widehat{\mathcal U}_{2\pi}^{(M)}
\boldsymbol\Xi_{\mathbf0}^{(M)}(-\pi),\\
\partial_\alpha\boldsymbol\Xi_{\mathbf0}^{(M)}(\pi)
&=\widehat{\mathcal U}_{2\pi}^{(M)}
\partial_\alpha\boldsymbol\Xi_{\mathbf0}^{(M)}(-\pi).
\end{aligned}
\label{eq:endmatter-global-vacuum-gluing}
\end{align}

The global ordinary and semiton states are constructed in the same way.
We define
\begin{align}
\boldsymbol\Phi_s^{(a)}(\alpha;g)
&:={}_{\rm rot}\langle\alpha|
\Psi_s^{(a)}[g]\rangle,
\qquad s\in\{0,1/2\}.
\label{eq:endmatter-global-packet-seed}
\end{align}
The state has $\mathbf F^{\rm can} = \mathbf e_a$.
By summing over the spectral-flow orbit, we obtain the $\theta=0$ states as
\begin{align}
\begin{aligned}
\boldsymbol\Psi_{s,\theta=0}^{(a)}(\alpha;g)
&:=\sum_{k\in\mathbb Z}
\bigl(\widehat{\mathcal U}_{2\pi}^{(M)}\bigr)^k
\boldsymbol\Phi_s^{(a)}(\alpha-2\pi k;g),
\\
|\Psi_{s,\theta=0}^{(a)}[g]\rangle
&:=\int_{-\pi}^{\pi}d\alpha\,
|\alpha\rangle_{\rm rot}\otimes
\boldsymbol\Psi_{s,\theta=0}^{(a)}(\alpha;g).
\end{aligned}
\label{eq:endmatter-global-packet-state}
\end{align}
The $k$-th summand now belongs to
$\mathbf e_a+k\mathbf q$.  Consequently,
\begin{align}
\int_{-\pi}^{\pi}d\alpha\,
\|\boldsymbol\Psi_{s,\theta=0}^{(a)}(\alpha;g)\|^2
&=\int_{\mathbb R}d\widetilde\alpha\,
\|\boldsymbol\Phi_s^{(a)}(\widetilde\alpha;g)\|^2=1.
\label{eq:endmatter-global-packet-normalization}
\end{align}
The state $\boldsymbol\Psi_{s,\theta=0}^{(a)}(\alpha;g)$ satisfies the gluing conditions in Eq.~\eqref{eq:endmatter-global-vacuum-gluing} for both $s=0$ and $s=1/2$.
The state is therefore a superposition of states with charges $F_i^{\rm can}=\delta_{ia}+k$.

Finally, orthogonality between different zero-mode sectors reduces its global ordinary--semiton overlap exactly to the branch-local one:
\begin{align}
\langle\Psi_{0,\theta=0}^{(a)}[g]|
\Psi_{1/2,\theta=0}^{(a)}[g]\rangle
&=\int_{\mathbb R}d\widetilde\alpha\,
\langle\boldsymbol\Phi_0^{(a)}(\widetilde\alpha;g)|
\boldsymbol\Phi_{1/2}^{(a)}(\widetilde\alpha;g)\rangle.
\label{eq:endmatter-global-overlap-reduction}
\end{align}

\subsection{Bosonized-vertex convention}

\newcommand{\vdslabel}{\text{\scriptsize[\citenum{vonDelftSchoeller:1998bosonization}]}}

The vertex $\mathcal V_{a,M}^{\dagger}$ in Eq.~\eqref{eq:ordinary-outgoing-wavepacket-state} is the sharp-cutoff version of the bosonization identity in Ref.~\cite{vonDelftSchoeller:1998bosonization}.
Taking the Hermitian conjugate of Eq.~(63) of Ref.~\cite{vonDelftSchoeller:1998bosonization} and using $[\widehat N_\eta,F_\eta^\dagger]=F_\eta^\dagger$, we obtain
\begin{align}
\begin{aligned}
\psi_\eta^\dagger\!\left(x_{\vdslabel}\right)
&=F_\eta^\dagger a_{\rm uv}^{-1/2}\\
&\quad\times\exp\!\left[i\frac{2\pi}{L_{\vdslabel}}
x_{\vdslabel}
\left(\widehat N_\eta+1-\frac{\delta_b}{2}\right)\right]\\
&\quad\times\exp\!\left[i\phi_\eta\!\left(
x_{\vdslabel}\right)\right],
\end{aligned}
\label{eq:endmatter-vds-creation-identity}
\end{align}
where $\phi_\eta$ is defined in Eq.~(34) of that reference.

The correspondence with the conventions of this Letter is
\begin{align}
\begin{aligned}
L_{\vdslabel}&=2L,
&\delta_b&=1,
\\
x_{\vdslabel}&=-x,
&\widehat N_a&=F_a^{\rm can},\\
q_n&=p_n,
&i\left.b_{q_n,a}\right|_{\vdslabel}&=b_{a,n},
\end{aligned}
\label{eq:endmatter-vertex-convention-map}
\end{align}
where the coordinate reversal accounts for the opposite chiral orientation
in the unfolded convention.  
We also replace the smooth short-distance cutoff by the sharp truncation $1\leq n\leq M$, and absorb the normalization into $\mathcal Z_M$. 
The Klein factor $F_a^\dagger$ is realized by
\begin{align}
F_a^\dagger|\mathbf n\rangle
&=(-1)^{\sum_{j<a}n_j}
|\mathbf n+\mathbf e_a\rangle,
\label{eq:endmatter-klein-cocycle-map}
\end{align}
where $F_a^\dagger$ commutes with $b_{i,n}^\dagger$ and $b_{i,n}$.
Then, Eqs.~\eqref{eq:endmatter-vds-creation-identity} and \eqref{eq:endmatter-vertex-convention-map} give
\begin{align}
\begin{aligned}
\mathcal V_{a,M}^{\dagger}(x)
&=\mathcal Z_M\,F_a^\dagger
\exp\!\left[-i\frac{\pi x}{L}
\left(F_a^{\rm can}+\frac{1}{2}\right)\right]\\
&\quad\times\exp\!\left[
\sum_{n=1}^{M}\frac{1}{\sqrt n}
\left(e^{-ip_nx}b_{a,n}^{\dagger}
-e^{ip_nx}b_{a,n}\right)\right].
\end{aligned}
\label{eq:endmatter-explicit-regulated-vertex}
\end{align}

\end{document}


\title{Supplemental Material for ``A Rotor-Dressed Semiton State in a Monopole--Fermion Model''}
\author{Yuta Hamada}
\affiliation{Theory Center, IPNS, High Energy Accelerator Research Organization (KEK), 1-1 Oho, Tsukuba, Ibaraki 305-0801, Japan}
\affiliation{Graduate Institute for Advanced Studies, SOKENDAI, 1-1 Oho, Tsukuba, Ibaraki 305-0801, Japan}
\affiliation{RIKEN Center for Interdisciplinary Theoretical and Mathematical Sciences (iTHEMS), RIKEN, Wako 351-0198, Japan}
\maketitle

\section{S1. From the Bosonized Boundary Action to the Current Hamiltonian}
\label{s1.-from-the-bosonized-boundary-action-to-the-current-hamiltonian}

We derive Eq.~(1) of the Letter from Eq.~(2.1) of Ref. [10] of the Letter.
Among $N=4$ rotors in Ref. [10], we only keep a relevant single rotor $\beta$ coupled to flavor-singlet boson $\Phi_{\Sigma}:=\sum_{i=1}^{4}\Phi_i/2$.
By denoting the inverse-inertia parameter of the rotor by $\mathcal I$, the bosonized half-line Lagrangian is
\begin{align}
L_{\rm Y}
={}&\frac{\dot\beta^2}{2\mathcal I}
+\frac{\beta}{2}\sum_{i=1}^{4}\dot\Phi_i(0,t)
+\frac1{8\pi}\sum_i\int_0^Ldx\,
\left(\dot\Phi_i^2-\Phi_i'^2\right),
\tag{S1a}
\end{align}
where the dot and prime denote time and spatial derivatives, respectively.
Starting with this action, we regularize the boundary coupling by a half-line core profile $f_+(x)$,
\begin{align}
\dot\Phi_i(0,t)&\longrightarrow
\int_0^Ldx\,f_+(x)\dot\Phi_i(x,t),
&\int_0^{r_0}dx\,f_+(x)&=1.
\tag{S1b}
\end{align}
The canonical momenta and the corresponding velocities are
\begin{align}
\Pi_\beta^{\rm Y}
&=\frac{\dot\beta}{\mathcal I},
&P_i(x)
&=\frac{\dot\Phi_i(x)}{4\pi}+\frac{\beta}{2} f_+(x).
\tag{S1c}
\end{align}
Consequently, the Legendre transform gives the Hamiltonian:
\begin{align}
H_{\rm Y}
=\Pi_\beta^{\rm Y}\dot\beta+\sum_i\int_0^Ldx\,P_i\dot\Phi_i-L_{\rm Y}
&=\frac{\mathcal I}{2}(\Pi_\beta^{\rm Y})^2
+\pi\sum_i\int_0^Ldx
\left\{
\left(P_i+\frac{\Phi_i^\prime}{4\pi}-\frac{\beta}{2} f_+\right)^2
+\left(P_i-\frac{\Phi_i^\prime}{4\pi}-\frac{\beta}{2} f_+\right)^2\right\}\nonumber\\
&=\frac{\mathcal I}{2}(\Pi_\beta^{\rm Y})^2
+\pi\sum_i\int_0^Ldx
\left\{\left(\rho_{i,+}-\frac{\beta}{2} f_+\right)^2+\left(\rho_{i,-}-\frac{\beta}{2} f_+\right)^2\right\}
\tag{S1d}
\end{align}
where we have introduced $\rho_{i,+}:=P_i+\frac{\Phi_i'}{4\pi}$ and $\rho_{i,-}:=P_i-\frac{\Phi_i'}{4\pi}$.
These are the chiral currents since $(\partial_t\mp\partial_x)\rho_{i,\pm}=0$ outside the core.

We now unfold the two chiral currents to a single current on $[-L,L]$ and define the core profile and the variables used in the Letter:
\begin{align}
f(x)&:=\frac{1}{2} f_+(|x|),
&\int_{-r_0}^{r_0}dx\,f(x)&=1,
&\alpha&:=2\pi\beta,
&\Pi&:=\frac{\Pi_\beta^{\rm Y}}{2\pi},
&I&:=\frac1{4\pi^2\mathcal I}.
\tag{S1e}
\end{align}
It follows that
$\beta f_+(|x|)/2=\beta f(x)=\alpha f(x)/(2\pi)$ and
$\mathcal I(\Pi_\beta^{\rm Y})^2/2=\Pi^2/(2I)$.  
Hence we have
\begin{align}
H_{\rm cur}
=\frac{\Pi^2}{2I}
+\pi\sum_{i}\int_{-L}^{L}dx\,
{:}\left(\rho_i^{\rm can}(x)
-\frac{\alpha}{2\pi}f(x)\right)^2{:},
\tag{S1f}
\end{align}
where $\rho_i^{\rm can}(x)$ is $\rho_{i,-}(x)$ for $x>0$ and $\rho_{i,+}(-x)$ for $x<0$.
After regularizing (S1f) to the first $M$ nonzero Fourier modes, it becomes Eq.~(1) of the Letter.

\section{S2. Overlap under the Point-excitation approximation}
\label{s2.-exact-overlap-and-the-finite-m-weyl-identity}

The purpose here is to derive Eq. (29) of the Letter.
The overlap between the ordinary and semiton states is
\begin{align}
\begin{aligned}
\langle\Psi_{0}^{(a)}[g]|\Psi_{1/2}^{(b)}[g]\rangle
=\frac{1}{\mathcal N}
\int dx\,dx'\,g(x)g(x')^*
\langle\Omega_{\mathbf0}^{(M)}|
\mathcal V_{a,M}(x')
\mathcal G^{(M)}[h]
\mathcal V_{b,M}^{\dagger}(x)
|\Omega_{\mathbf0}^{(M)}\rangle
=:\delta^{ab}\langle\Psi_{0}|\Psi_{1/2}\rangle.
\end{aligned}
\tag{S2a}
\end{align}
Here $\mathcal N$ is the wavepacket-vertex normalization defined in the Letter.

The \emph{point-excitation approximation} is defined by replacing both the smeared fermion excitation and its density profile by their regularized point counterparts:
\begin{align}
\mathcal V_{a,M}^{\dagger}[g]
\longmapsto
\frac{1}{|\mathcal Z_M|}\mathcal V_{a,M}^{\dagger}(X),
\qquad
|g(x)|^2\longmapsto\delta_M(x-X),
\tag{S2b}
\end{align}
where $\delta_M$ is defined in Eq.~(15) of the Letter, and $\mathcal Z_M$ is defined by
\begin{align}
\mathcal V_{a,M}(X)\mathcal V_{a,M}^{\dagger}(X)
=|\mathcal Z_M|^2\mathbf1.
\tag{S2c}
\end{align}
The profile $h$ is also replaced by the regularized point profile
\begin{align}
h_{\rm pt}(x):=f(x)-\delta_M(x-X).
\tag{S2d}
\end{align}
Within the point-excitation approximation, we make the replacement
\begin{align}
\left|\langle\Psi_{0}|\Psi_{1/2}\rangle\right|
\to\left|\langle\Psi_{0}^{\rm pt}|\Psi_{1/2}^{\rm pt}\rangle\right|
\tag{S2e}
\end{align}
where
\begin{align}
&|\Psi_{0}^{\rm pt}\rangle
:=\frac{1}{|\mathcal Z_M|}
\mathcal V_{a,M}^{\dagger}(X)|\Omega_{\mathbf0}^{(M)}\rangle,
&&|\Psi_{1/2}^{\rm pt}\rangle
:=\frac{1}{|\mathcal Z_M|}
\mathcal G^{(M)}[h_{\rm pt}]
\mathcal V_{a,M}^{\dagger}(X)|\Omega_{\mathbf0}^{(M)}\rangle.
\tag{S2f}
\end{align}


Using the algebraic relations, we obtain
\begin{align}
\mathcal V_{a,M}(X)\mathcal G^{(M)}[h_{\rm pt}]
\mathcal V_{a,M}^{\dagger}(X)
=|\mathcal Z_M|^2
\exp\!\left[-i\sum_{n=1}^{M}
\frac{\operatorname{Im}(h_{{\rm pt},n}e^{-ip_nX})}{n}\right]
\mathcal G^{(M)}[h_{\rm pt}],
\tag{S2g}
\end{align}
from which we have
\begin{align}
\begin{aligned}
\left|\langle\Psi_{0}^{\rm pt}|\Psi_{1/2}^{\rm pt}\rangle\right|
=\left|\langle\Omega_{\mathbf0}^{(M)}|
\mathcal G^{(M)}[h_{\rm pt}]
|\Omega_{\mathbf0}^{(M)}\rangle\right| .
\end{aligned}
\tag{S2h}
\end{align}
As we can see from Eqs.~(6) and (18) in the Letter, $\mathcal G$ can be written as
\begin{align}
&\mathcal G^{(M)}[h_{\rm pt}] = e^{i(\Delta\boldsymbol\pi^Tz-\Delta z^T\boldsymbol\pi)},
&&\Delta z=\left(\pi,\sqrt{\frac{2L}{\pi}}\frac{\mathrm{Re}h_{{\rm pt},n}}{n}\right),
&&\Delta\boldsymbol\pi=\left(0,-\sqrt{\frac{2\pi}{L}}\mathrm{Im}h_{{\rm pt},n}\right).
\tag{S2i}
\end{align}
Therefore, we need to compute the vacuum expectation value
\begin{align}
\begin{aligned}
&\left|
\left\langle\Omega_{\mathbf0}^{(M)}\right|
e^{i(\Delta\boldsymbol\pi^Tz-\Delta z^T\boldsymbol\pi)}
\left|\Omega_{\mathbf0}^{(M)}\right\rangle
\right|
=\exp\!\left[-\frac14\Delta z^TK\Delta z
-\frac14\Delta\boldsymbol\pi^TK^{-1}\Delta\boldsymbol\pi\right]
\leq 
\exp\!\left[-\frac14\Delta\boldsymbol\pi^TK^{-1}\Delta\boldsymbol\pi\right],
\end{aligned}
\tag{S2j}\label{eq:s2-gaussian-weyl-expectation}
\end{align}
The last inequality follows from the positive definiteness of $K$.
Next, we compute the kernel $(K^{-1})_{xx}$ using the fact $\langle\Omega_{\mathbf0}^{(M)}|\mathbf{x} \mathbf{x}^T |\Omega_{\mathbf0}^{(M)}\rangle=(K^{-1})_{xx}/2$, which we compute from the quadratic part $x$ in the action.
The Euclidean action of the rotor-current system is (see Eq. (7) in the Letter for the Hamiltonian)
\begin{align}
S_E
&=\frac12\int_\omega\!\left\{
(I \omega^2+C)\alpha_{-\omega}\alpha_{\omega}
-\alpha_{-\omega}\bm c^T\bm x_{\omega}
-\bm x_{-\omega}^T\bm c\,\alpha_{\omega}
+\bm x_{-\omega}^T(\omega^2\mathbf1_M+D^2)\bm x_{\omega}
\right\}
\nonumber\\
&=\frac12\int_\omega
(I\omega^2+C)
\left[\alpha_{-\omega}
-\frac{\bm x_{-\omega}^T\bm c}{I\omega^2+C}\right]
\left[\alpha_{\omega}
-\frac{\bm c^T\bm x_{\omega}}{I\omega^2+C}\right]
+\frac12\int_\omega\bm x_{-\omega}^T
\left[\omega^2\mathbf1_M+D^2-\frac{\bm c\bm c^T}{I\omega^2+C}\right]
\bm x_{\omega}.
\tag{S2k}
\label{eq:fixed-free-end-current-covariance-bound}
\end{align}
where $\omega$ is the Euclidean frequency, $\int_\omega:=\int_{-\infty}^{\infty}d\omega/(2\pi)$, and $D_{nm}=p_n\delta_{nm}$ with $n,m=1,\ldots,M$.
After integrating out the rotor, we see
\begin{align}
\frac12\bigl(K^{-1}\bigr)_{xx}
=\int_\omega
\left[\omega^2\mathbf1_M+D^2-\frac{\bm c\bm c^T}{I\omega^2+C}\right]^{-1}
\succeq
\int_\omega
\left[\omega^2\mathbf1_M+D^2\right]^{-1}
=\int_\omega\frac{\delta_{nm}}{\omega^2+p_n^2}
=\frac{{\delta_{nm}}}{2p_n},
\tag{S2l}
\end{align}
where $\succeq$ means the inequality of matrices. For example, $A\succeq B$ means $A-B$ is positive semidefinite.
Substituting this into Eq.~(S2j) gives the bound
\begin{align}
\left|\langle\Psi_{0}^{\rm pt}|\Psi_{1/2}^{\rm pt}\rangle\right|
\leq
\exp\!\left[-\sum_n\frac{1}{4}\frac{\Delta\pi_n^2}{p_n}\right]
=\exp\!\left[-\sum_n\frac{\left(\mathrm{Im}h_{{\rm pt},n}\right)^2}{2n}\right]
\tag{S2m}
\end{align}

Combining this with the Fourier transform of the point profile in Eq.~(S2d),
\begin{align}
&h_{{\rm pt},n}
=f_n-e^{ip_nX},
&&\operatorname{Im}h_{{\rm pt},n}
=-\sin(p_nX)=-\sin\frac{\pi nX}{L},
\tag{S2n}\label{eq:s2-point-profile-fourier-mode}
\end{align}
the bound becomes
\begin{align}
\begin{aligned}
\left|\langle\Psi_{0}^{\rm pt}|\Psi_{1/2}^{\rm pt}\rangle\right|
\leq\exp\!\left[-\frac{1}{2}\sum_{n=1}^{M}\frac{\sin^2(\pi n R/L)}{n}\right]
&\to\exp\!\left[-\frac{1}{2}\int_0^{R\Lambda} dp\frac{\sin^2(p)}{p}\right]
\simeq \text{const.}\times(R\Lambda)^{-1/4},
\end{aligned}
\tag{S2o}\label{eq:s2-harmonic-window-logarithm}
\end{align}
where we take $L\to\infty$ in the step represented by the arrow, and then $R\Lambda\to\infty$ in the last step.
This confirms Eq. (29) of the Letter.

\section{S3. Retarded Ward derivation of the half-unit core density}
\label{s3.-retarded-ward-derivation-of-the-half-unit-core-density}

We now derive Eq. (17) of the Letter.
Using the current definition in Eq. (3) of the Letter, its
finite-core Heisenberg equations are
\begin{align}
&(\partial_t+\partial_x)J_i^{\rm cov}(x,t)
=-\frac{f(x)}{2\pi I}\Pi(t),
&&\dot\Pi(t)
=\sum_{j=1}^{4}\int_{-r_0}^{r_0}dx\,
f(x)J_j^{\rm cov}(x,t),
&&\dot\alpha(t)=\frac{\Pi(t)}{I}.
\tag{S3a}
\end{align}

Integration of the transport equation in (S3a) gives
\begin{align}
J_i^{\rm cov}(x,t)=J_i^{\rm cov}(-r_0,t-x-r_0)-\frac{1}{2\pi I}\int^x_{-r_0}dx^\prime f(x^\prime)\Pi(t-x+x^\prime).
\tag{S3b}
\end{align}
For $|x|\leq r_0$ and in the low-energy regime ($Er_0\ll1,\,IE\ll1$), we can approximate the above as
\begin{align}
J_i^{\rm cov}(x,t)
&\simeq J_i^{\rm cov}(-r_0,t)
-\frac{\Pi(t)}{2\pi I}\int_{-r_0}^{x}dy\,f(y).
\tag{S3c}
\end{align}
Substituting this into the rotor equation, we obtain
\begin{align}
\dot\Pi(t)+\frac{1}{\pi I}\Pi(t)
&\simeq J_{\Sigma}^{\rm in}(t),
&
J_{\Sigma}^{\rm in}(t)
&:=\sum_{j=1}^{4}J_j^{\rm cov}(-r_0,t) .
\tag{S3d}
\end{align}
Let us consider a first-scattering time window, where the rotor is not excited at the initial and end times.
For one normalized incoming fermion, $\int dt\,\langle J_{\Sigma}^{\rm in}(t)\rangle=1$, by integrating (S3d) over this window, we obtain
\begin{align}
    \frac{1}{\pi}\int dt \frac{\langle\Pi\rangle}{I}\simeq \int dt \langle J_{\Sigma}^{\rm in}(t)\rangle=1.
\tag{S3e}
\end{align}
Using $\dot\alpha=\Pi/I$, the above equation gives
\begin{align}
\Delta\langle\alpha\rangle
\simeq\pi .
\tag{S3f}
\end{align}
Here $\Delta\alpha$ is the response on the chosen local lift of the compact rotor.

From (S3b), after the first scattering and for $|x|\leq r_0$, $\langle J_i^{\rm cov}(x,t)\rangle$ vanishes as there are no incoming currents, and the rotor term decays exponentially.
Combining with (S3f) yields Eq.~(17) of the Letter.
